\documentclass[%
  reprint,
  amsmath,amssymb,
 aps,
]{revtex4-2}

\usepackage{graphicx}
\usepackage{dcolumn}
\usepackage{bm}
\usepackage{xcolor} 
\begin{document}

\preprint{APS/123-QED}

\title{Tests of gravitational wave propagation with LIGO-Virgo catalog}

\author{Xian-Liang Wang$^{1,2,3}$}

\author{Shu-Cheng Yang$^{4}$}%

\author{Wen-Biao Han$^{4,1,3}$}%
 \email{wbhan@shao.ac.cn}
\affiliation{
   $^1$School of Fundamental Physics and Mathematical Sciences, Hangzhou Institute for Advanced Study,~UCAS, Hangzhou 310024, China \\
   $^2$Institute of Theoretical Physics,~UCAS, Beijing 100190, China\\
   $^3$University of Chinese Academy of Sciences, 100190 Beijing, China\\
   $^4$Shanghai Astronomical Observatory, Shanghai 200030, China\\
   }

\date{\today}

\begin{abstract}
In the framework of general relativity (GR), gravitational waves (GWs) travel at the speed of light across all frequencies. However, massive gravity and weak equivalence principle (WEP) violation may lead to frequency-dependent variations in the propagation speed of GWs, which can be examined by comparing the theoretical and observed discrepancies in the arrival times of GW signals at various frequencies. This provides us with an opportunity to test these theories. For massive gravity, we consider that gravitons may have a nonzero rest mass. For WEP violations, we hypothesize that different massless particles exposed to the same gravitational source should exhibit varying gravitational time delays. The gravitational time delay induced by massive gravitational sources is proportional to $\gamma+1$, where the parameter $\gamma=1$ in GR. Therefore, we can quantify these two deviations using phenomenological parameters $m_g$ and $|\Delta \gamma|$, respectively. In this study, we use selected GW data from binary black hole coalescences in the LIGO-Virgo catalogs GWTC-2.1 and GWTC-3 to place constraints on the parameters $m_g$ and $|\Delta \gamma|$. We analyze the relationship between $m_g$ and luminosity distance,as well as between $|\Delta \gamma|$ and both luminosity distance sky location of GW events to determine the presence of graviton mass and WEP violation. Nevertheless, we find no evidence of such relationships. We also compute Bayes factors for models that assume the existence of graviton mass and WEP violation compared to the standard GW model, respectively. The absolute value of the natural logarithm of the Bayes factor is generally less than 2. Our analysis reveals no significant preference for either model. Additionally, the Bayes factors between these two models do not provide obvious evidence in favor of either one.
 
\end{abstract}

\maketitle

\section{\label{sec:level1}Introduction}
The detection of gravitational waves (GWs) by LIGO \cite{aasi2015advanced} and Virgo \cite{acernese2014advanced} collaborations heralds a new era in both astrophysics and fundamental physics \cite{abbott2016ligo,abbott2017gw170817,abbott2019gwtc,abbott2021gwtc2,abbott2024gwtc,abbott2023gwtc3}. During the third observing run (O3), Advanced LIGO, Advanced Virgo and KAGRA added 79 GW events, as detailed in GWTC-2.1 \cite{abbott2024gwtc} and GWTC-3 \cite{abbott2023gwtc3}. These discoveries deepen our understanding of the universe and provide powerful tools to check the rules of gravitational theories such as general relativity (GR), and help us better understand the basic structure of spacetime.

According to GR, GWs are theorized to propagate at the speed of light in a vacuum, and their velocity $v_g$ is predicted to be invariant across all frequencies. This characteristic suggests that the propagation speed of GWs is unaffected by frequency, ensuring uniformity in the velocity across different wavelengths. The GWs detected by Advanced LIGO and Advanced Virgo can be used to make statistical inferences about $v_g$\cite{cornish2017bounding,liu2020measuring,ray2023measuring,Ghosh:2023xes}, and the speed of GWs can be constrained within the range of $0.99^{+0.01}_{-0.02}c$ at a 90\% credible interval\cite{ray2023measuring}. In the framework of the quantum gravity theory, gravitons are postulated to be carriers for transmitting gravitational interactions. GWs travel at the speed of light across all frequencies which supports the idea that gravitons are massless.

However, certain modified gravitational theories typically anticipate violations of Lorentz invariance (LI) \cite{Amelino-Camelia:2002cqb,Amelino-Camelia:2010lsq,Sefiedgar:2010we,Horava:2008ih,Horava:2009uw,Garattini:2011es,Garattini:2011kp,garattini2012particle}, indicating that the speed of GWs might not always align with the speed of light and could vary based on their frequencies during propagation \cite{hovrava2009quantum,amelino2010doubly,sefiedgar2011modified,garattini2011modified}. These theories imply that gravitons may possess a rest mass that is not equal to zero. We use modified dispersion relations to depict these theories\cite{mirshekari2012constraining},and we focus our discussion on a specific scenario, which is the massive gravity\cite{will1998bounding}. It is hypothesized that low-frequency GWs propagate slower than high-frequency GWs. Consequently, during the inspiral phase of compact binary stars, gravitons emitted earlier have a lower propagation speed than those emitted later. This phenomenon induces phase alterations in GWs during propagation, resulting in received waveforms that differ from those predicted by GR. If such dephasing of waveforms comparing with GR's templates is not detected, this could imply a constraint on the graviton mass, thereby limiting the potentially extent of LI violation. The graviton mass has been measured through various experimental datasets\cite{shoom2022constraining,abbott2019tests,abbott2021tests,abbott2021tests3,wu2023search,wu2023constraining}.

Similar to the impact of the massive gravity, if the weak equivalence principle (WEP) is violated, the speed of GWs will no longer be the speed of light but will depend on the frequencies. This frequency dependence allows us to limit the extent of WEP violation. The violation of WEP results in different Shapiro delays \cite{shapiro1964fourth} for different massless particles traversing an identical gravitational field. The Shapiro delay is proportional to $\gamma + 1$, where $\gamma$ is the parametrized post-Newtonian parameter ($\gamma= 1$ in general relativity) \cite{will2018theory,will2014confrontation}, and different massless particles will have different values of $\gamma$ during free fall in a gravitational field if WEP is violated. It is often considered that two different massless particles are emitted from the same astrophysical source, with respective values of $\gamma_1$ and $\gamma_2$, then $|\Delta \gamma|$ ($|\gamma_{1} - \gamma_{2}|$) can be used to quantitatively represent the violation of the WEP. If we know the intrinsic time delay of these two particles, we can limit the difference in the Shapiro time delay between the two particles $\Delta t_{\rm gra}$ \cite{krauss1988test} which is proportional to $|\Delta \gamma|$, by measuring the observed time-delays $\Delta t_{\rm obs}$. Therefore, we can establish an upper limit for $|\Delta \gamma|$ by observing the $\Delta t_{\rm obs}$ of various signals emitted from the same source. Several constraints on $|\Delta \gamma|$ have been derived from various astrophysical events, including the emissions from the supernova event SN1987A \cite{krauss1988test,longo1988new}, gammy-ray bursts \cite{gao2015cosmic,sang2016testing,yu2018new,wei2020testing,yi2020constraining,bartlett2021constraints} and fast radio bursts (FRBs) \cite{hashimoto2021upper,wei2015testing,tingay2016limits,xing2019limits}. In this study, we use GWs to test WEP, which had previously been employed \cite{kahya2016constraints,wu2016testing,wei2017multimessenger,yang2020tests}. The utilization of GW signals can eliminate the intrinsic time delay when signals are emitted, which will enable us to estimate the up-limit of $|\Delta \gamma|$ more accurately.

Building on the previously discussed framework, we further explore the phenomena of massive gravity and WEP violation, both of them modify the speed of GWs resulting in frequency dependence. Specifically, if gravitons possess mass, the speed of gravitons emitted during the early inspiral phase of GW emission is expected to be lower than that emitted later. Consequently, the impact of this effect on the GW waveform is expected to become more pronounced as the luminosity distance of GW events from Earth increases. This is consistent with the behavior under WEP violation, where the influence of gravitational fields leads to cumulative effects as the luminosity distance of the GW increases. However, it is essential to note that WEP violation manifests through changes in the speed of GWs due to the influence of gravitational fields.

In this study, our analysis focuses on the gravitational potential generated by the Milky Way, under the assumption that its entire mass is concentrated at the center. Owing to the unclear details of the host galaxies and galaxies along the propagation paths of these events, we neglect the time delays caused by these galaxies. This approach renders the constraints in our study very conservative, as the impacts from the host galaxy and other galaxies may possess similar or even greater magnitude compared to those from the Milky way. Therefore, the actual values of $|\Delta \gamma|$ might be several times smaller than our results.

The remainder of this paper is organized as follows. Section 2 introduces our methodology for analyzing the specific dispersion of GWs. In Section 3, we present our results for 22 GW events. Conclusions and discussions are presented in Section 4.

\section{Method}
We use the common phenomenological modification to GR introduced in  \cite{mirshekari2012constraining}, which was previously applied to LIGO and Virgo data in  \cite{scientific2017gw170104,abbott2019tests,abbott2021tests,abbott2021tests3}:
\begin{eqnarray}
E^2=p^2c^2+A_\alpha p^\alpha c^\alpha,
\label{one}
\end{eqnarray}
where c is the speed of light, E and p are the energy and momentum of the GWs, respectively, $A_\alpha$ and $\alpha$ are phenomenological parameters. With different $A_\alpha$ and $\alpha$, we can use Eq.~(\ref{one}) to denote different modified dispersion relations in different alternative gravitational theories. In this study, we only consider the scenarios where $\alpha=0$ and $A_0 > 0$, for which the graviton mass is given by $m_g=A_0^{1/2}/c^2$, following the framework of massive gravity\cite{will1998bounding}.

\begin{figure*}
\includegraphics[width=1\textwidth]{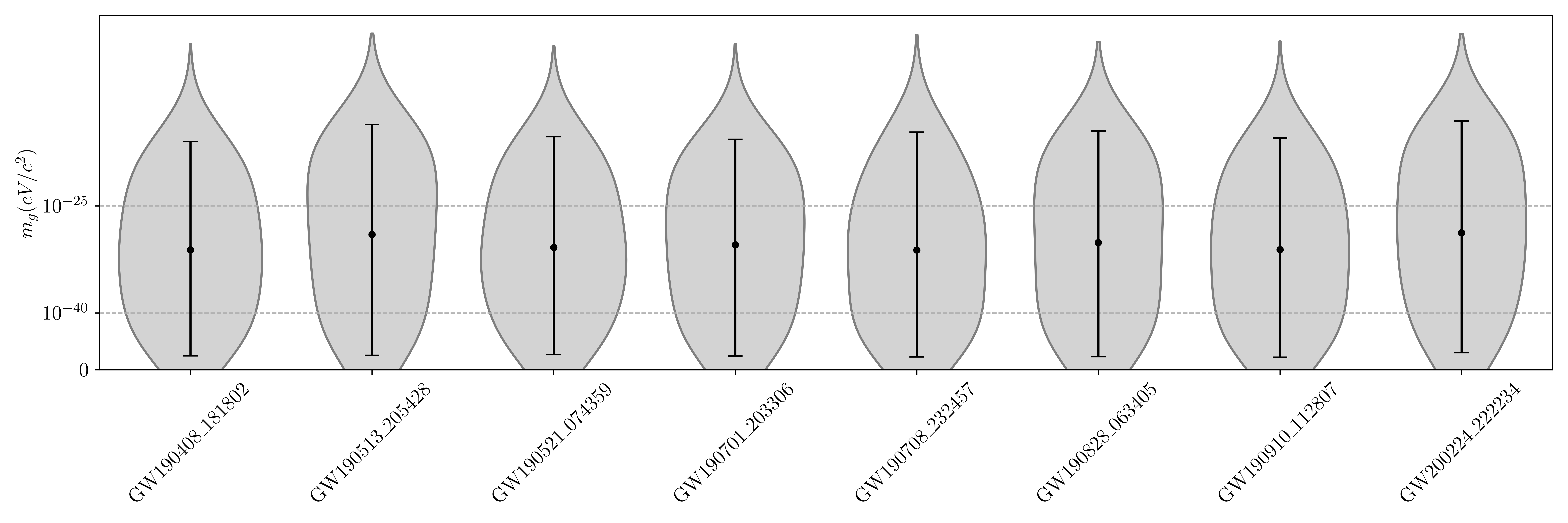}
\caption{\label{fig:mgwide}The posterior distribution of $m_g$ for selected GW events of BBH in GWTC-2.1 and GWTC-3 (90\% credible interval).}
\end{figure*}
\begin{figure*}
\includegraphics[width=1\textwidth]{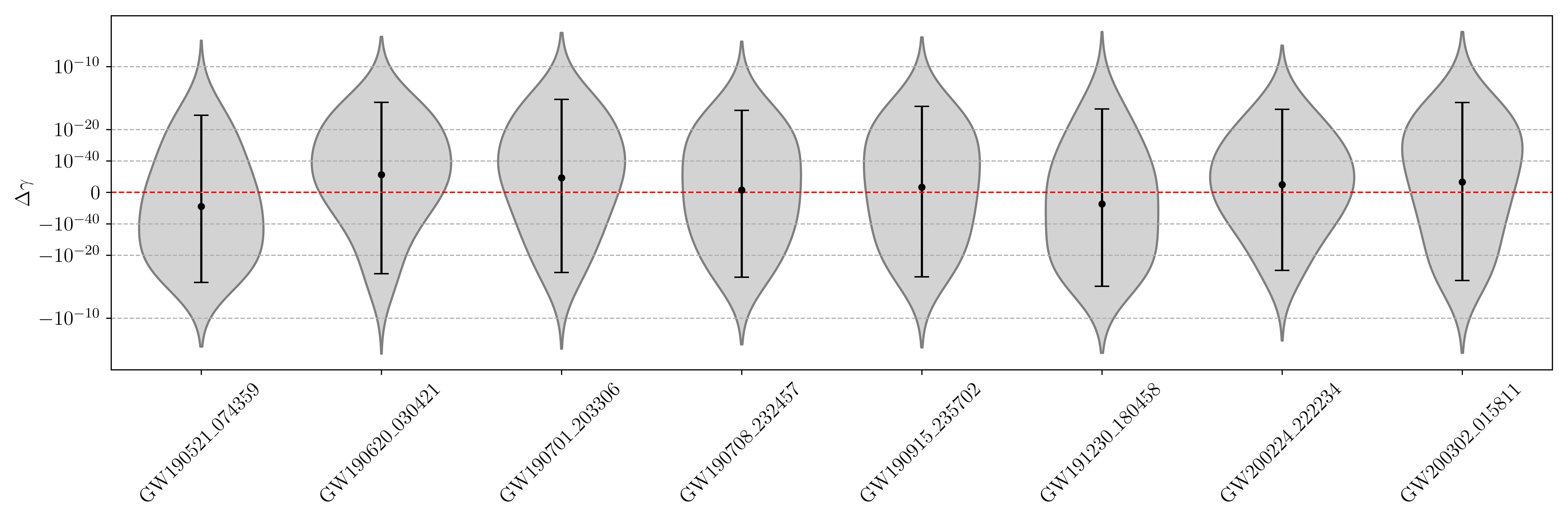}
\caption{\label{fig:gammawide}The posterior distribution of $\Delta \gamma$ for selected GW events of BBH in GWTC-2.1 and GWTC-3 (90\% credible interval). We selected a modified logarithmic prior (refer to Eqs.(6) in \cite{yang2020tests}) for the prior distribution of $\Delta \gamma$, which encompasses both negative and non-negative values.}
\end{figure*}

The modified gravitational theory will lead to a difference in the dispersion relation of GWs, which will lead to the propagation of GWs at a speed different from that of light. Therefore, a modified waveform template that includes such effects is generally used to describe the waveform of GWs, which is expressed in the frequency domain as follows

\begin{eqnarray}
\tilde{h}(f)=\tilde{A}(f)\mathrm{e}^{i[\Psi_{\mathrm{GR}}(f)+\delta\Psi(f)]},
\label{two}
\end{eqnarray}
where $\tilde{A}(f)$ denotes the complex amplitude, $\Psi_{\mathrm{GR}}(f)$ denotes the complex phase predicted by GR, and $\delta\Psi(f)$ is the modification term. In the scenarios where $\alpha=0$ and $A_0 > 0$, it is given by 
\begin{equation}
\delta\Psi_{\alpha=0}(f)=-\frac{{\pi}D_0{A_0}}{h^2c(1+z)f},
\end{equation}
where $z$ is the cosmological redshift, $h$ is the Planck constant, $f$ is the frequency of the GWs, and $D_0$ is the distance parameter given by
\begin{equation}
D_0=\frac{c(1+z)}{H_0}\int_0^z\frac{(1+\tilde{z})^{-2}d\tilde{z}}{\sqrt{\Omega_M(1+\tilde{z})^3+\Omega_\Lambda}},
\end{equation}
where $H_0 = 67.90 \text{ km s}^{-1} \text{ Mpc}^{-1}$ is the Hubble constant, $\Omega_\text{m} = 0.3065$ and $\Omega_\Lambda = 0.6935$ are the matter and dark energy density parameters, respectively. These are the TT+lowP+lensing+ext values from \cite{ade2016planck}.

The Shapiro time delays generated during the propagation of GWs will also induce frequency-dependent variations in their speed owing to the violation of WEP. Here, we assume that $\delta\Psi(f)$ is caused by WEP deviation. Considering a GW event, GWs emitted at $t_{\rm{e}}$ and $t'_{\rm{e}}$ with different frequencies are received by the detectors at the corresponding arrival times $t_{\rm{a}}$ and $t'_{\rm{a}}$. We can ignore the cosmological inflation effect if the difference in emitting time ($\Delta t_{\rm{e}} = t_{\rm{e}} - t'_{\rm{e}}$) is so small that the difference in the arrival times of the corresponding GWs ($\Delta t_{\rm{a}} = t_{\rm{a}} - t'_{\rm{a}}$) is
\begin{eqnarray}
\Delta t_\mathrm{a}=\left(1+z\right)[\Delta t_\mathrm{e}+\Delta t_\mathrm{gra}],
\label{ge}
\end{eqnarray}
In this study, we only consider the violation of the WEP caused by the Milky Way, and $\Delta t_{\rm gra}$  would be \cite{wu2016testing}
\begin{equation}
\begin{aligned}
\Delta t_{\mathrm{gra}} =&\Delta\gamma\left[\frac{GM_{\mathrm{MW}}}{c^{3}}\right.  \\
&\left.\times\mathrm{ln}\left(\frac{\left[d+(d^2-b^2)^{1/2}\right]\left[r_G+s_n\left(r_G^2-b^2\right)^{1/2}\right]}{b^2}\right)\right],
\end{aligned}
\label{six}
\end{equation}
where the Milky Way mass $M_{\mathrm{MW}} \approx 6 \times 10^{11}~M_{\odot}$ and the distance from the sun to the center of the Milky Way $r_{G} \approx 8 ~ \mathrm{kpc}$, $d$ denotes the distance from the GW event to the Milky Way center, $b$ represents the impact parameter of the GW paths relative to the center of the Milky Way. We use a transform formula \cite{yao2020testing} to convert the celestial coordinates to $b$. $s_n = + 1$ denotes the GW event positioned in the direction of the Milky Way center, and  $s_n = -1$ denotes the GW event positioned away from the Milky Way center. However, Minazzoli et al. have raised concerns about the application of Eq.~(\ref{six}) for calculating Shapiro delay on cosmological scales \cite{Minazzoli:2019ugi}, as these concerns extend beyond the current study's scope, we exclude them from our analysis. Then, we obtain $\delta\Psi(f)$
\begin{equation}
\begin{aligned}
\delta\Psi(f) =&\frac{\pi\Delta\gamma(1+z)f^{2}}{\Delta f}\left[\frac{GM_{\mathrm{MW}}}{c^{3}}\right.  \\
&\left.\times\mathrm{ln}\left(\frac{\left[d+(d^2-b^2)^{1/2}\right]\left[r_G+s_n\left(r_G^2-b^2\right)^{1/2}\right]}{b^2}\right)\right],
\end{aligned}
\end{equation}
where $\Delta f = f - f'$ and $f, ~f'$ are two different frequencies of GWs in a single event. It is intuitive to assume that $\gamma$ is proportional to the particle energy. Considering $E = h f$, we assume that $\Delta \gamma \propto \Delta f$. 

\begin{figure}[t]
\includegraphics[width=0.5\textwidth]{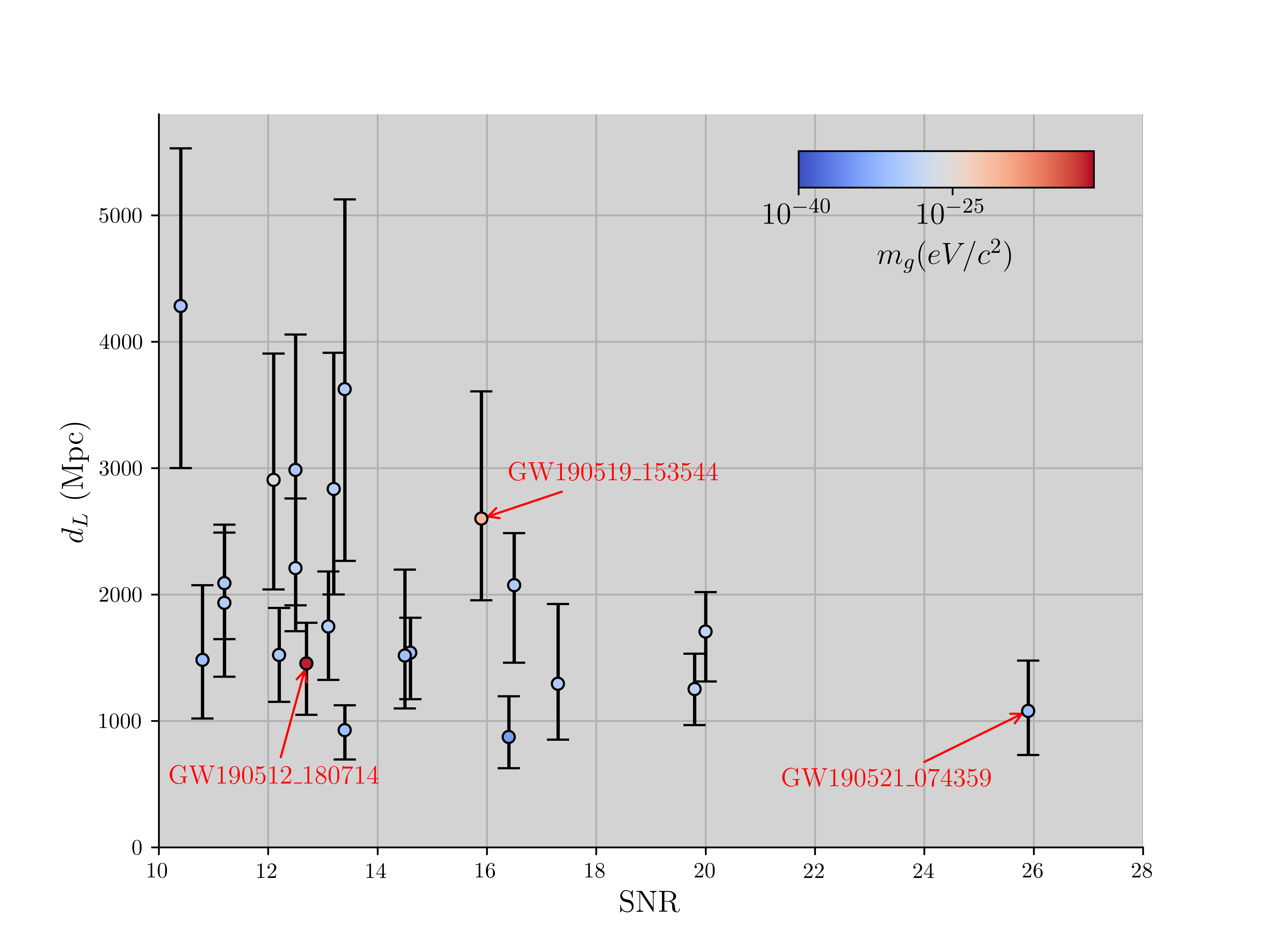}
\caption{\label{fig:dlcolor} The median of the posterior distribution of $m_g$ for selected GW events of BBH in GWTC-2.1 and GWTC-3, represented by the color of points. Results are displayed with 1$\sigma$ credible interval for the luminosity distance.}
\end{figure}
\begin{figure}[t]
\includegraphics[width=0.5\textwidth]{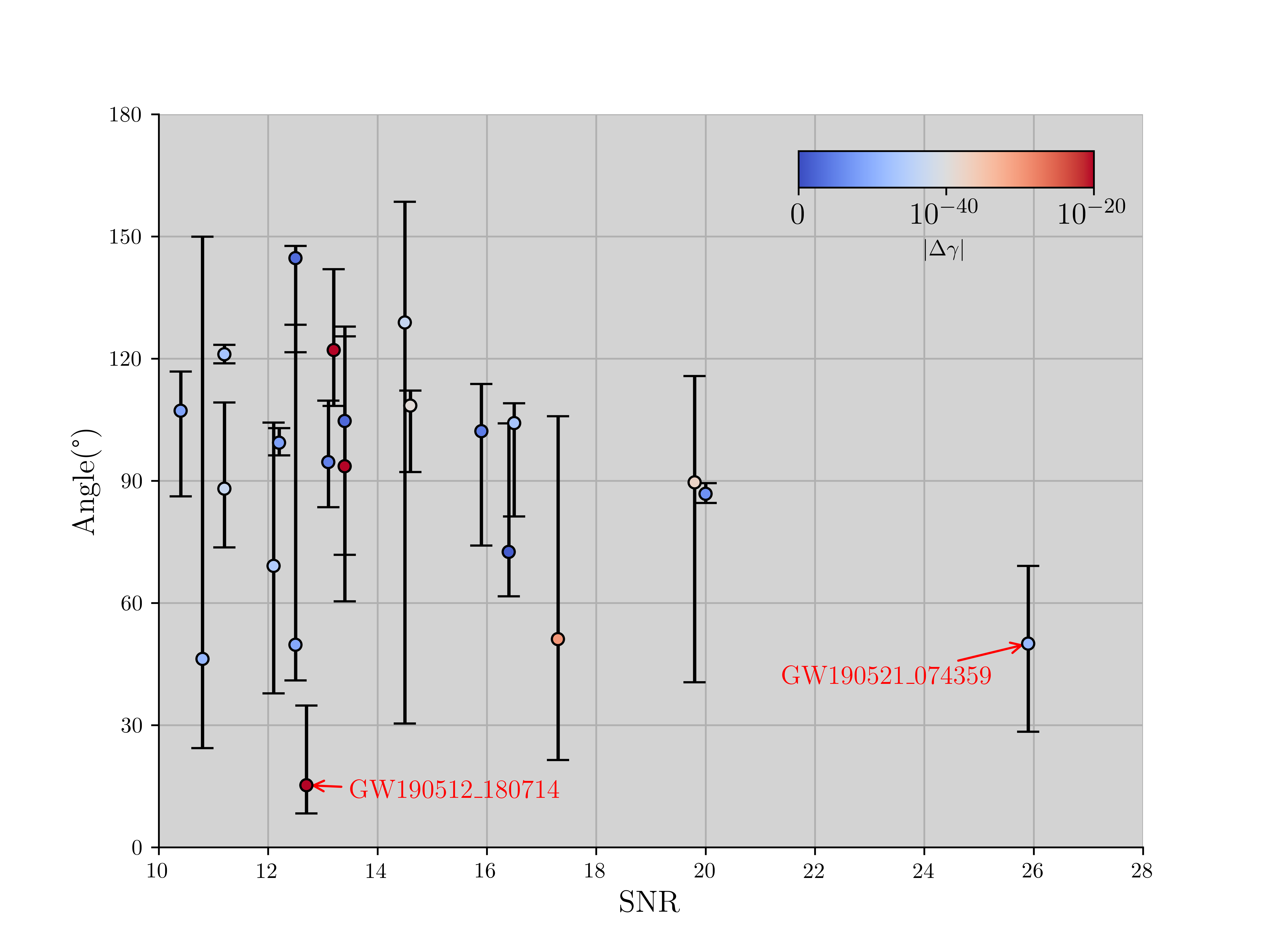}
\caption{\label{fig:dgcolor} The median of the posterior distribution of $\Delta\gamma$ for selected GW events of BBH in GWTC-2.1 and GWTC-3, represented by the color of points. Results are displayed with 1$\sigma$ credible interval for the angle.}
\end{figure}

In this study, we employ the BILBY Bayesian parameter estimation software \cite{ashton2019bilby} and the DYNESTY nested sampling package \cite{Speagle:2019ivv} to perform parameter estimation. We use the parametrized waveform model described in Eq.~(\ref{two}), which incorporates modification terms into the IMRPhenomXP waveforms \cite{Pratten:2020ceb}. These modifications are designed to estimate parameters $m_g$ and $\Delta\gamma$ using the selected GW events in GWTC-2.1 and GWTC-3. The GW strain data are obtained from the Gravitational Wave Open Science Center (GWOSC), which is publicly accessible at [https://gwosc.org/]. For the parameter priors, we adopt BILBY’s default settings for binary black holes (BBHs) \cite{ashton2019bilby} except for $m_g$ and $\Delta\gamma$. For $\Delta \gamma$, we select a modified logarithmic prior that encompasses both negative and non-negative values (refer to Eqs.(6) in \cite{yang2020tests}). Furthermore, inspired by this approach, we similarly define the prior distribution for $m_g$ as
\begin{equation}
{m_g}(\alpha) = \left\{
\begin{aligned}
&10^{-\frac{1}{\alpha}}, && \text{for } \alpha > 0 \\
&0, && \text{for } \alpha = 0
\end{aligned}
\right.
\label{pri}
\end{equation}
where $\alpha$ is a uniform distribution parameter. Based on previous studies, we set the prior range for $m_g$ as [0, $10^{-20}eV/c^2$] and for $\Delta\gamma$ as [$-10^{-7}$, $10^{-7}$].

If deviations are present, indicating non-zero values for $m_g$ or $\Delta\gamma$. Specifically, in the context of massive gravity, we hypothesize that an increase in the luminosity distance of the GW event correlates with a more pronounced impact on the GW signal. For WEP violation, we hypothesize that the impact on the GW signal intensifies as the propagation path of the GWs approaches closer to the center of the Milky Way, and also increases with increasing luminosity distance of the GW event. 

However, the further the luminosity distance of a GW event, the lower its Signal-to-Noise Ratio (SNR) tends to be. In the presence of the deviations, discrepancies from the standard GW waveforms become more obvious with increasing luminosity distance for both models. But the data precision of these GW events decreases correspondingly. Therefore, we can selectively analyze events with similar SNRs but significant differences in luminosity distance to examine their estimated values and error ranges more accurately. If the deviation exist, it is anticipated that events with greater luminosity distances and similar SNRs will result in more precise estimations of $m_g$ and $\Delta\gamma$, with smaller margins of error.

We also can ignore the influence of SNR and use the Bayes factor to directly compare the advantages and disadvantages of the modified models and the standard model using real GW data. 
\begin{equation}
\mathrm{BF}_{B}^{A}=\frac{\mathcal{Z}_{A}}{\mathcal{Z}_{B}},
\end{equation}
where $\mathcal{Z}_{A}$,$\mathcal{Z}_{B}$ is the Bayesian evidence of models A and B. To enhance the data’s intuitiveness, we use the natural logarithm of the Bayes factor.
\begin{equation}
\ln\mathrm{BF}_{B}^{A}=\ln(\mathcal{Z}_A)-\ln(\mathcal{Z}_B).
\end{equation}

\begin{figure*}
\includegraphics[width=1\textwidth]{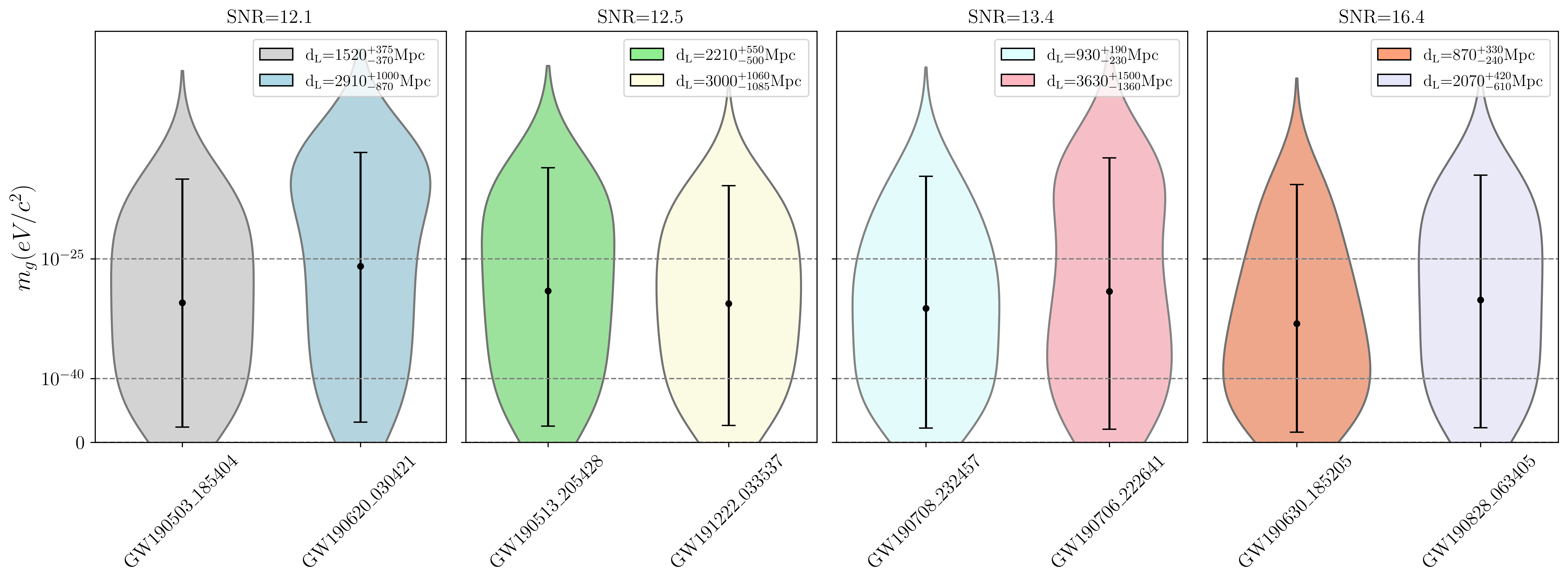}
\caption{\label{fig:mg4setwide}The posterior distribution of $m_g$ for selected GW events of BBH in GWTC-2.1 and GWTC-3 (90\% credible interval). The groups are arranged from left to right in increasing order of SNR, with events on the left having a smaller luminosity distance than those on the right in each group}
\end{figure*}

\begin{figure*}
\includegraphics[width=1\textwidth]{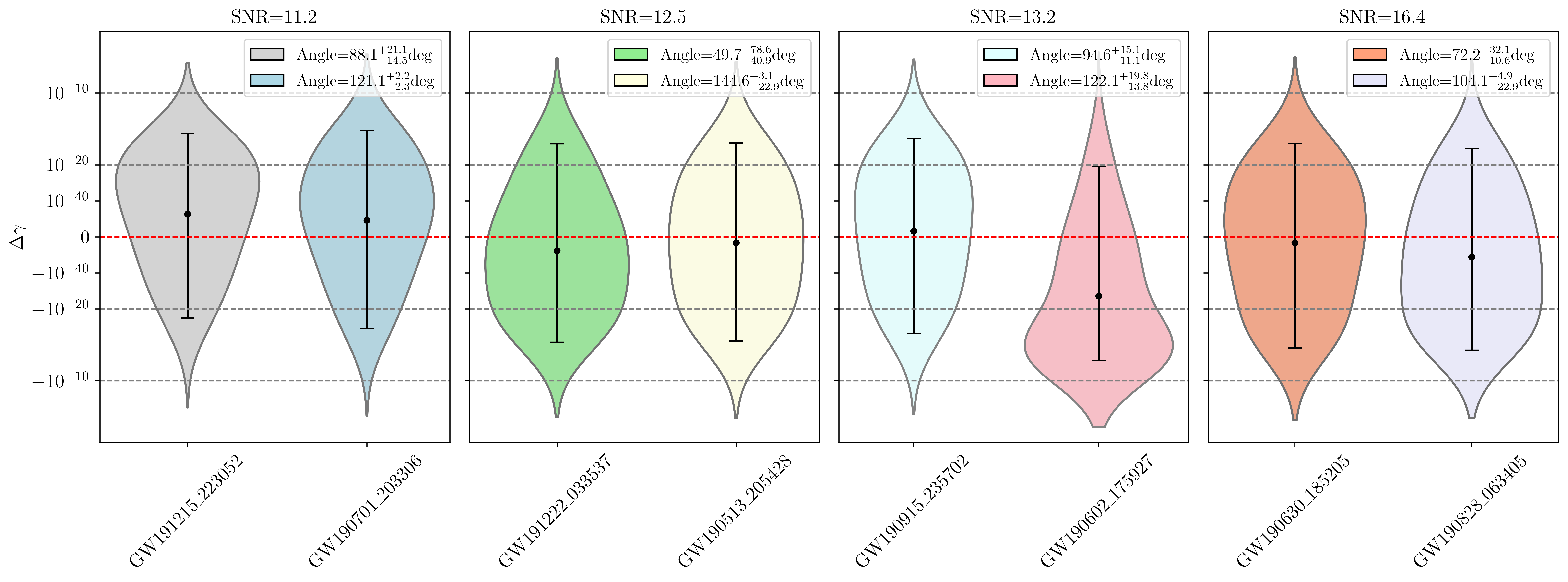}
\caption{\label{fig:gamma4setwide}The posterior distribution of $\Delta\gamma$ for selected GW events of BBH in GWTC-2.1 and GWTC-3 (90\% credible interval). The groups are arranged from left to right in increasing order of SNR, with events on the left having a smaller angle than those on the right in each group.}
\end{figure*}

Massive gravity and the violation of WEP induce frequency-dependent variations in the propagation speed of GWs. Assuming that the impact of GW waveforms caused by WEP violation is solely attributable to the Milky Way, it follows that as the angle between the GW event and the Milky Way center as observed from Earth gradually decreases, its impact should become more pronounced. Conversely, the impact of GW waveforms caused by massive gravity should be independent of the angles of GW events and would uniformly affect the propagation of GWs through spacetime across all directions. This impact is expected to exhibit cumulative effects over the propagation distance. In other words, as the propagation time of GW increases, so does the luminosity distance of the GW events, the deviation of GW waveforms will become increasingly evident. We can elucidate this phenomenon using Eq.~(\ref{two}), where a larger luminosity distance corresponds to a greater variation in $\delta\Psi$. This variation enables the differentiation of the GW waveform from the one predicted by GR. Thus, this induces an increasing impact on the GW waveform as the luminosity distance of the GW event increases. 

Thus, as the luminosity distance of GW events increases, the Bayes factor between the massive gravity and the standard model derived from GWs data will increase. We also hypothesize that the potential impact of the WEP violation would be magnified as the luminosity distance of GW events increases and the angle decreases. Consequently, the Bayes factor comparing WEP violation model to the standard model based on GW data should exhibit higher values, indicating more pronounced WEP violation effects. Conversely, lower values indicate weaker effects.

\begin{figure}[b]
\includegraphics[width=0.5\textwidth]{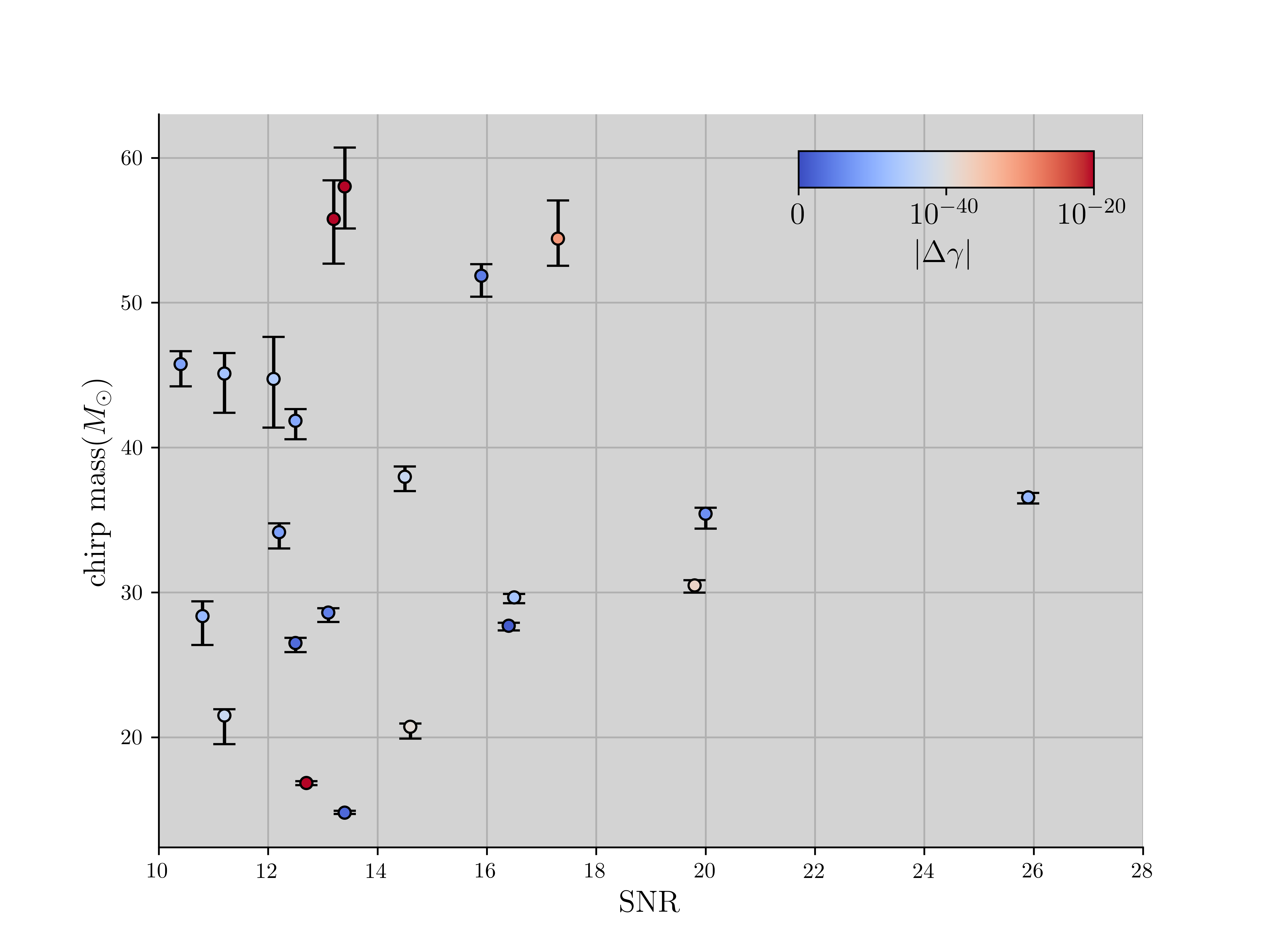}
\caption{\label{figother}The median of the posterior distribution of $\Delta\gamma$ for selected GW events of BBH in GWTC-2.1 and GWTC-3, represented by the color of points. Results are displayed with 1$\sigma$ credible interval for the chirp mass}
\end{figure}

\section{Results}
Our data is based on GW events of BBH in GWTC-2.1 and GWTC-3. We select GW events that exhibit a SNR greater than 10. We calculate the angle between each GW event and the center of the Milky Way as observed from Earth. Using the median of the posterior distribution of angles as the reference value, we approximately uniformly select GW events across the entire range of angles. In this selection, events within 0 to 30 degrees and 150 to 180 degrees are relatively rare. For events with similar angles, we select the one with higher SNR to do the analysis. Following this methodology, we then obtain 22 suitable GW events.

\begin{figure}[ht]
\includegraphics[width=0.5\textwidth]{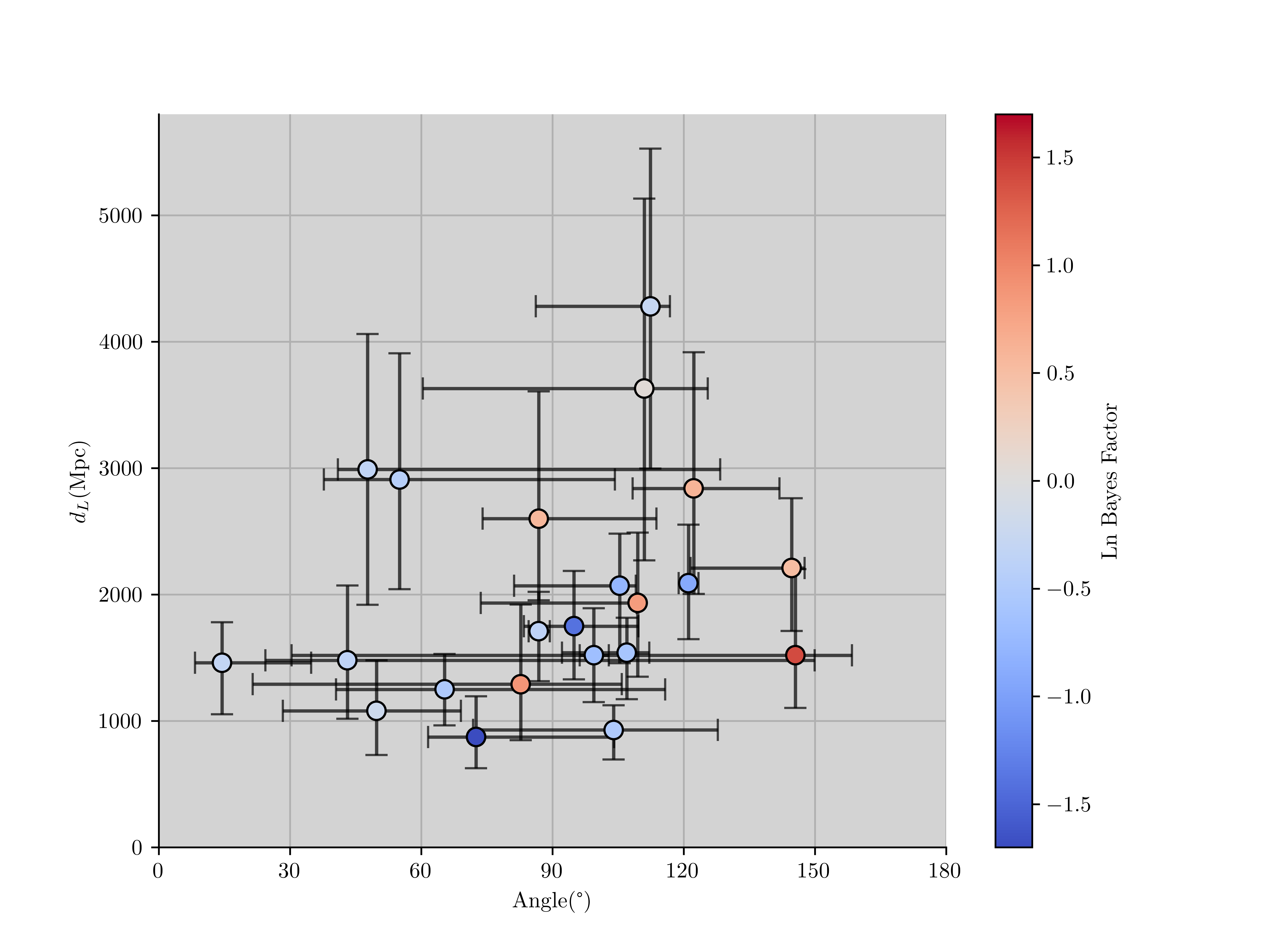}
\caption{\label{figtwo}The natural logarithm of the Bayes factors of WEP violation model versus standard model for selected GW events of BBH in GWTC-2.1 and GWTC-3, represented by the color of points. Results are displayed with 1$\sigma$ credible interval for the angle and luminosity distance.}
\end{figure}

Fig.~\ref{fig:mgwide} and Fig.~\ref{fig:gammawide} illustrate the posterior distributions of $m_g$ and $\Delta\gamma$ for a subset of events (90\% credible interval), respectively. The prior distribution of $m_g$ follows the distribution presented in Eq.~(\ref{pri}). The prior distribution of $\Delta\gamma$ follows the distribution presented in Eqs.(6) in \cite{yang2020tests}, which provides continuous coverage for both negative and non-negative $\Delta\gamma$ value. The detailed posterior information of $\Delta \gamma$ and $m_g$ for each event is tabulated in Table ~\ref{table1}. Furthermore, the potential data-quality issues inherent in GW observations have been discussed in studies\cite{abbott2024gwtc,abbott2023gwtc3}. It is crucial to acknowledge that our method could not measure the exact values of $m_g$ and $|\Delta \gamma|$. This limitation arises from the assumption that all observation errors are attributable to the deviation arise from massive gravity and WEP violation. Consequently, the results presented here represent an upper limit for $m_g$ and $|\Delta \gamma|$, implying that the actual values of $m_g$ and $|\Delta \gamma|$ should be lower than the estimations obtained in this study.

\begin{figure}[b]
\includegraphics[width=0.5\textwidth]{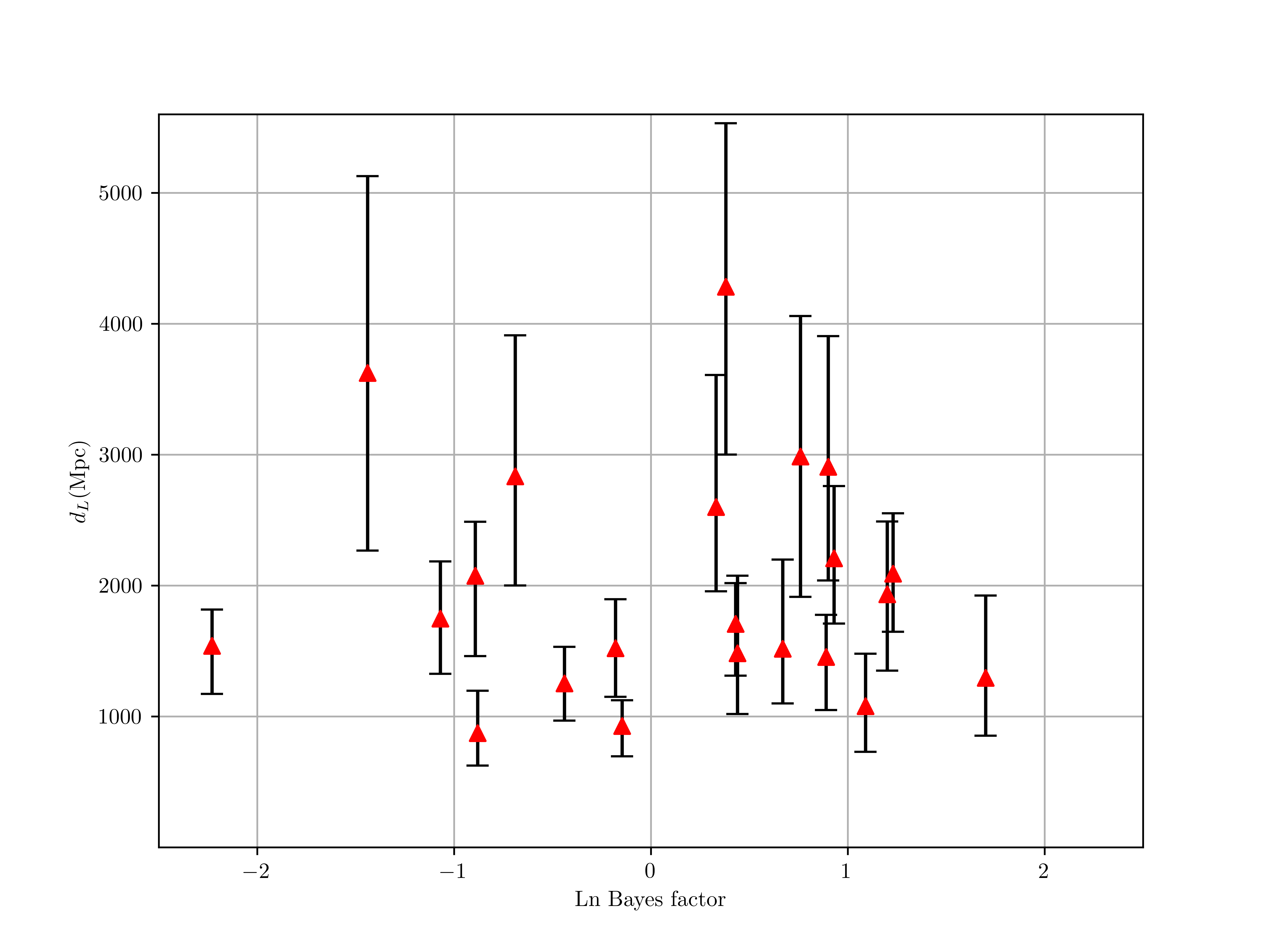}
\caption{\label{fig:three} The natural logarithm of the Bayes factors of massive gravity versus standard model for selected GW events of BBH in GWTC-2.1 and GWTC-3. Results are displayed with 1$\sigma$ credible interval for the luminosity distance.}
\end{figure}

\begin{figure}[t]
\includegraphics[width=0.5\textwidth]{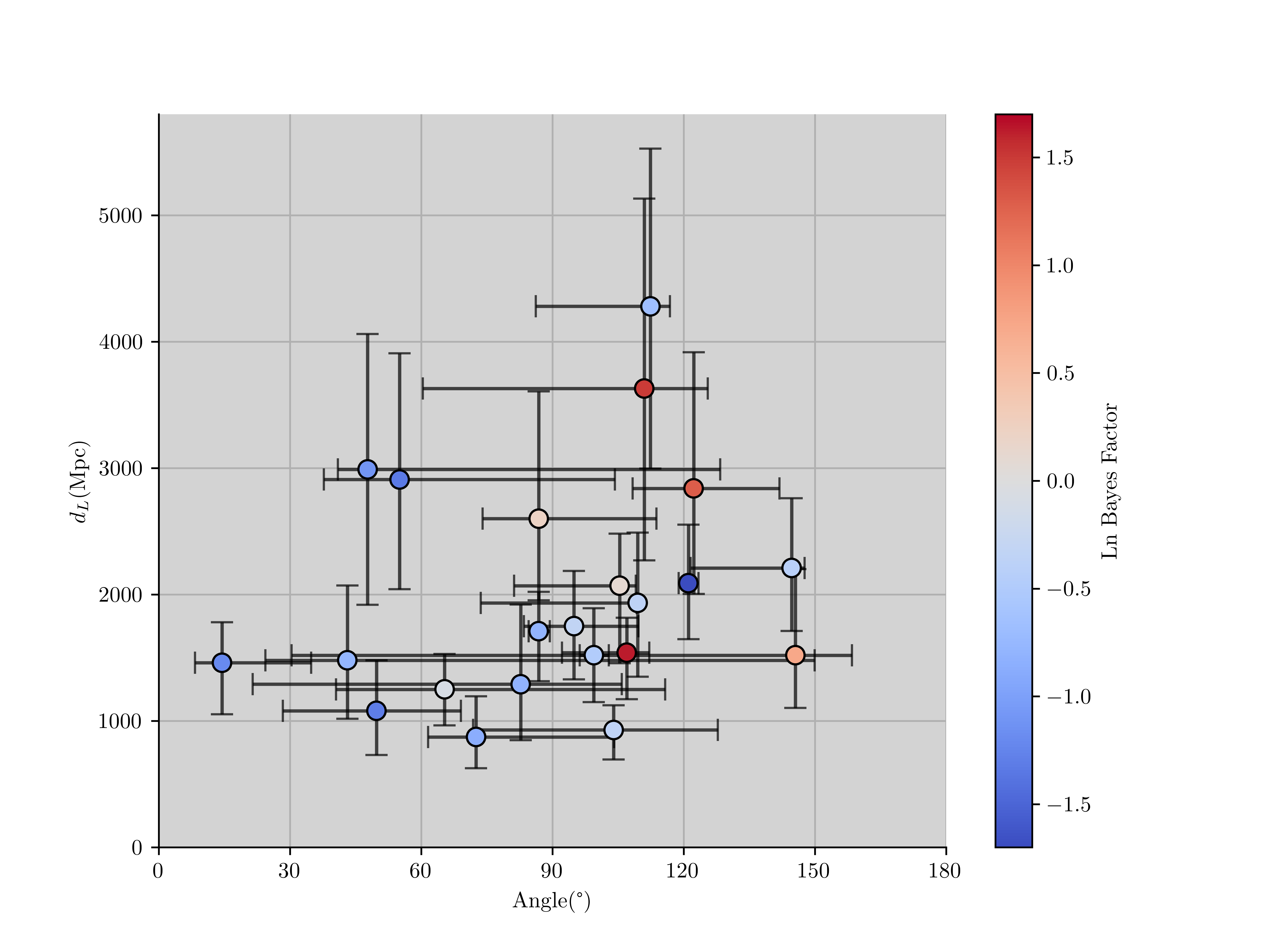}
\caption{\label{fig:four} The natural logarithm of the Bayes factors of WEP violation model versus massive gravity for selected GW events of BBH in GWTC-2.1 and GWTC-3, represented by the color of points. Results are displayed with 1$\sigma$ credible interval for the angle and luminosity distance.}
\end{figure}

In Fig.~\ref{fig:dlcolor}, contrary to expectations, GW events at a greater luminosity distance and higher SNR cannot provide more tighter constraints on $m_g$ than other events. Specifically, high-SNR event such as GW190521\_074359, the observed outcome cannot demonstrate superior results in terms of estimated $m_g$ compared to other low-SNR events. In cases with similar SNR, events with greater luminosity distances, such as GW190519\_153544 and GW190512\_180714, their outcomes still fail to demonstrate superior results in terms of estimated $m_g$ compared to other events. 

In Fig.~\ref{fig:dgcolor}, we do not observe the anticipated trend where GW events with smaller angles and higher SNRs would exhibit tighter constraints on $\Delta\gamma$ than other events. Specifically, high-SNR event GW190521\_074359, its outcome cannot demonstrate superior results in terms of estimated $\Delta\gamma$ compared to other events. In cases with similar SNR, GW190512\_180714 with small angle, the outcome still fails to demonstrate superior results in terms of the estimated $\Delta\gamma$ compared to other events.

For the analysis of massive gravity, we select four groups of GW events with similar SNRs differing by only 0.1, but exhibiting substantial variations in luminosity distance as shown in Fig.~\ref{fig:mg4setwide}. The GW events are arranged from left to right in increasing order of SNR, with events on the left having a smaller luminosity distance than those on the right in each group. Due to the logarithmic scale used for the coordinates of $m_g$, it is essential to focus on the upper limit of the 90\% credible interval of the error bars, rather than their total length. We observed that at the SNR of 12.5, the upper limit for the $m_g$ at the 90\% credible interval of GW191222\_033537 at a luminosity distance of $3000^{+1700}_{-1700}$ Mpc is tighter than that of GW190513\_205428 at $2210^{+550}_{-500}$ Mpc. The differences in the upper limit for the $m_g$ at the 90\% credible interval for other groups do not show significant variability.

\begin{table*}

\caption{\label{table1}The posterior distribution of $\Delta \gamma$ within the 90\% credible interval and 90\% credible interval upper bounds on the graviton mass $m_g$ for the selected BBH GW events in GWTC-2.1 and GWTC-3.}
\begin{ruledtabular}
\renewcommand{\arraystretch}{1.7} 
\begin{tabular}{ccc} 
 Events & Posterior distribution of $\Delta\gamma$ &   $m_g$[$eV/c^2$]\\ \hline
 
        GW190408\_181802 & ${-1.96\times10^{-13}}\leq\Delta\gamma\leq{6.20\times10^{-17}}$    &${1.79\times10^{-23}}$ \\
        GW190503\_185404 & ${-6.69\times10^{-16}}\leq\Delta\gamma\leq{2.27\times10^{-13}}$    &${2.62\times10^{-23}}$ \\
        GW190512\_180714 & ${-3.89\times10^{-13}}\leq\Delta\gamma\leq{1.19\times10^{-24}}$    &${1.37\times10^{-22}}$ \\
        GW190513\_205428 & ${-1.48\times10^{-14}}\leq\Delta\gamma\leq{4.80\times10^{-16}}$    &${4.49\times10^{-23}}$ \\
        GW190519\_153544 & ${-5.83\times10^{-13}}\leq\Delta\gamma\leq{2.96\times10^{-16}}$    &${7.90\times10^{-23}}$ \\
        GW190521\_074359 & ${-1.08\times10^{-14}}\leq\Delta\gamma\leq{4.86\times10^{-17}}$    &${2.37\times10^{-23}}$ \\
        GW190602\_175927 & ${-2.28\times10^{-12}}\leq\Delta\gamma\leq{2.92\times10^{-21}}$    &${4.61\times10^{-23}}$ \\
        GW190620\_030421 & ${-3.49\times10^{-16}}\leq\Delta\gamma\leq{1.03\times10^{-14}}$    &${8.59\times10^{-23}}$ \\
        GW190630\_185205 & ${-1.06\times10^{-13}}\leq\Delta\gamma\leq{3.75\times10^{-16}}$    &${2.02\times10^{-23}}$ \\
        GW190701\_203306 & ${-2.00\times10^{-16}}\leq\Delta\gamma\leq{2.93\times10^{-14}}$    &${2.01\times10^{-23}}$ \\
        GW190706\_222641 & ${-3.72\times10^{-12}}\leq\Delta\gamma\leq{8.61\times10^{-18}}$    &${6.91\times10^{-23}}$ \\
        GW190708\_232457 & ${-1.56\times10^{-15}}\leq\Delta\gamma\leq{4.74\times10^{-16}}$    &${3.02\times10^{-23}}$ \\
        GW190828\_063405 & ${-2.04\times10^{-13}}\leq\Delta\gamma\leq{3.82\times10^{-20}}$    &${3.15\times10^{-23}}$\\
        GW190910\_112807 & ${-3.39\times10^{-13}}\leq\Delta\gamma\leq{2.24\times10^{-16}}$    &${2.17\times10^{-23}}$ \\
        GW190915\_235702 & ${-1.24\times10^{-15}}\leq\Delta\gamma\leq{2.39\times10^{-15}}$    &${2.17\times10^{-23}}$ \\
        GW191109\_010717 & ${-1.82\times10^{-13}}\leq\Delta\gamma\leq{2.52\times10^{-22}}$    &${1.55\times10^{-23}}$ \\
        GW191215\_223052 & ${-1.85\times10^{-18}}\leq\Delta\gamma\leq{1.18\times10^{-14}}$    &${2.66\times10^{-23}}$ \\
        GW191222\_033537 & ${-2.27\times10^{-14}}\leq\Delta\gamma\leq{3.49\times10^{-16}}$    &${1.90\times10^{-23}}$ \\
        GW191230\_180458 & ${-4.10\times10^{-14}}\leq\Delta\gamma\leq{8.26\times10^{-16}}$    &${1.68\times10^{-23}}$ \\
        GW200112\_155838 & ${-2.54\times10^{-13}}\leq\Delta\gamma\leq{2.47\times10^{-18}}$    &${6.91\times10^{-23}}$ \\
        GW200224\_222234 & ${-7.88\times10^{-17}}\leq\Delta\gamma\leq{7.05\times10^{-16}}$    &${5.38\times10^{-23}}$ \\
        GW200302\_015811 & ${-5.42\times10^{-15}}\leq\Delta\gamma\leq{9.87\times10^{-15}}$    &${2.55\times10^{-23}}$ \\  
\end{tabular}
\end{ruledtabular}
\end{table*}

For the analysis of WEP violation, we selected four groups of GW events with similar SNRs differing by only 0.1, but exhibiting substantial variation in the angles between the GW event and the center of the Milky Way, as shown in Fig.~\ref{fig:gamma4setwide}. The GW events are arranged from left to right in increasing order of SNR, with events on the left having a smaller angle than those on the right in each group. Here, it is important to consider both the upper and lower limits of the 90\% credible interval of the error bars.  At the SNR = 13.2, GW190915\_235702 with an angle of ${94.6^{+15.1}_{-11.1}}$ degrees establishes a tighter limit at the 90\% credible interval on the $|\Delta\gamma|$ than the GW event GW190602\_175927 with an angle of ${122.1^{+19.8}_{-13.8}}$ degrees. The variations in the upper limit for the $|\Delta\gamma|$ at the 90\% credible interval for the remaining groups do not show significant variability.

In fact, potential degeneracies between the deviations from massive gravity or WEP violation and other source parameters could introduce imprecision in the estimated values of $m_g$ and $\Delta\gamma$. However, we present upper limits, the results are considered acceptable. Fig.~\ref{figother} presents the relationship between the chirp mass and $|\Delta\gamma|$ within our posterior distributions.

Additionally, we employ Bayes factors to evaluate the validity of the massive gravity and WEP violation models relative to the standard GW model. Fig.~\ref{figtwo} presents the natural logarithm of Bayes factors between the WEP violation model and the standard model for selected GW events of BBH in GWTC-2.1 and GWTC-3. The natural logarithm of Bayes factors are predominantly below an absolute value of 2 and do not show any significant trends in the top-left or bottom-right corners of the distribution, which correspond to the scenarios of small angle with large luminosity distance and large angle with small luminosity distance. These outcomes indicate that significant deviations in Bayes factors are not evident, suggesting a negligible impact of WEP violation on GW propagation under the current experimental conditions and data precision in our study.

Fig.~\ref{fig:three} presents the natural logarithm of Bayes factors between the massive gravity and the standard model for selected GW events of BBH in GWTC-2.1 and GWTC-3. Similar to the previous findings, the natural logarithm of Bayes factors are generally below an absolute value of 2, with no obvious deviation that correlate with variations in luminosity distance. These outcomes suggest the absence of significant deviations in Bayes factors relative to the luminosity distance under the current experimental conditions and data precision, which means that the massive gravity cannot provide a superior fit to the data compared to the standard model.

Further, we also calculate Bayes factors between the WEP violation model and massive gravity, as shown in Fig.~\ref{fig:four}. The Bayes Factors are below zero in approximately two-thirds of the events. However, the absolute values of these natural logarithm of Bayes factors are generally less than 2. These results indicate that there is insufficient evidence to prefer either model. Therefore, the GW data do not provide sufficient grounds to assert the superiority of one model over the other in terms of its fit to the GW observations.

\section{Discussion}
Since the direct detection of GWs \cite{abbott2016observation}, the LIGO and Virgo collaboration has rigorously evaluated the consistency between observed GW signals \cite{scientific2017gw170104,abbott2021tests,abbott2021tests3} and the theoretical predictions derived from GR. Therefore, previous studies have already placed bounds on the graviton mass and tested potential deviations induced by violation of the WEP during the propagation of GWs, and no significant deviations from the predictions of GR have been revealed through the examination of the GW data. It should be noted, however, that under the current experimental conditions, it is still necessary to use some novel methods to examine the existence of graviton mass and to test WEP violation during the propagation of GWs.

In this study, we do not find a significant relationship between the parameter $m_g$ and luminosity distance and a relationship between the parameter $|\Delta \gamma|$ and both luminosity distance and sky location of the GW events. We do not find a significant preference for either the massive gravity or WEP violation models. Moreover, Bayes factors fail to provide obvious evidence favoring one model over the other. With the current level of experimental precision, it remains uncertain to determine whether these deviations exist.

In this study, we have exclusively focused on the scenarios where $\alpha=0$ and $A_0 > 0$ in modified dispersion relations. In an upcoming work, we will consider more cases of $A_\alpha$ with different $\alpha$, and will adopt a more comprehensive theory that incorporates the effects of WEP and LI violations, which will allow us to directly compare the impact of these two effects on GWs. In the method of detecting the properties of GWs, we will no longer just focus on phase variations in the frequency domain of GW waveform. However, we will investigate how violations of WEP and LI may influence on other properties of GWs, such as their polarization.

The fourth observing run (O4) of the LIGO-Virgo-KAGRA (LVK) GW detector network has started running, it is promising to get more GW events, including more binary neutron star (BNS) merger events. These events are anticipated to have a promising probability of presenting multi-messenger characteristics, which would permit a better determination of parameters such as right ascension, declination, and luminosity distance. Then, we can clearly analyze the differences in propagation speeds between GWs and electromagnetic waves and the impact of large mass gravitational sources on the propagation of GWs. So it is possible to better constrain the upper bound of $m_g$ and $\Delta\gamma$. 

In the future, the implementation of space-borne interferometers will expand our ability to detect GWs from a wider array of sources and improve the SNRs of detected events \cite{danzmann2017proposal,amaro2017laser,armano2018beyond,hu2017taiji,luo2016tianqin}. We will also develop statistical methods and try more efficient ways to search for traces of deviations from GR in the multitude of GW events. The utilization of O4 data and forthcoming space-borne interferometers will allow us to detect various GW events, and conduct population analysis on GW events that may contain deviations from GR. Hence, it is worthwhile to discuss the capability of GW detectors to detect signals that exhibit such deviations in future studies.

\begin{acknowledgments}
This study was supported by the National Key R\&D Program of China (Grant No.2021YFC2203002), the National Natural Science Foundation of China (Grant No. 12173071). This study made use of the HPC Cluster of ITP-CAS and the High Performance Computing Resource in the Core Facility for Advanced Research Computing at Shanghai Astronomical Observatory.
\end{acknowledgments}

\end{document}